\documentclass{aa}  

\usepackage{graphicx}

\usepackage{txfonts,textcomp}
\usepackage{gensymb} 
\usepackage{xspace}
\usepackage{tabularx}
\usepackage{placeins}
\usepackage{threeparttable}

\usepackage{natbib,twoopt}
\usepackage[breaklinks=true]{hyperref} 
\bibpunct{(}{)}{;}{a}{}{,} 
\makeatletter
\newcommandtwoopt{\citeads}[3][][]{\href{http://adsabs.harvard.edu/abs/#3}%
{\def\hyper@linkstart##1##2{}%
\let\hyper@linkend\@empty\citealp[#1][#2]{#3}}}
\newcommandtwoopt{\citepads}[3][][]{\href{http://adsabs.harvard.edu/abs/#3}%
{\def\hyper@linkstart##1##2{}
\let\hyper@linkend\@empty\citep[#1][#2]{#3}}}
\newcommandtwoopt{\citetads}[3][][]{\href{http://adsabs.harvard.edu/abs/#3}%
{\def\hyper@linkstart##1##2{}
\let\hyper@linkend\@empty\citet[#1][#2]{#3}}}
\newcommandtwoopt{\citeyearads}[3][][]%
{\href{http://adsabs.harvard.edu/abs/#3}
{\def\hyper@linkstart##1##2{}%
\let\hyper@linkend\@empty\citeyear[#1][#2]{#3}}}
\makeatother
\def\m2s2{\hbox{\,m$^{2}$\,s$^{-2}$}} 
\def\Msun{$M_{\odot}$\xspace}             
\def\Rsun{$R_{\odot}$\xspace}

\def\ten[#1]{$\;\times 10^{#1}$}

\usepackage{hyperref}
\usepackage{xcolor}
\hypersetup{unicode}
\hypersetup{breaklinks=true}
\hypersetup{colorlinks,					
	linkcolor={red!50!black},
	citecolor={blue!50!black},
	urlcolor={blue!80!black}}

\newcommand{\astropy}{{\sc \tt astropy}\xspace}
\newcommand{\prose}{{\sc \tt prose}\xspace}
\newcommand{\photutils}{{\sc \tt photutils}\xspace}
\newcommand{\pytranspot}{{\sc \tt PyTranSpot}\xspace}
\newcommand{\juliet}{{\sc \tt juliet}\xspace}

\newcommand{\REarth}{R$_{\oplus}$\xspace}
\newcommand{\MEarth}{M$_{\oplus}$\xspace}

\def\Msun{$M_{\odot}$\xspace}            
\def\Rsun{$R_{\odot}$\xspace}
\newcommand{\orcid}[1]{\protect\href{https://orcid.org/#1}{\protect\includegraphics[width=8pt]{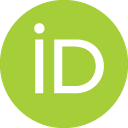}}}
\hypersetup{colorlinks=true, citecolor=blue, linkcolor=blue}

\begin{document}

   \title{The changing transit shape of TOI-3884 b}

    \author{
        H.~Chakraborty\thanks{\href{mailto:Hritam.Chakraborty@unige.ch}{Corresponding author: Hritam.Chakraborty@unige.ch}}\orcid{0000-0002-5177-1898}\inst{\ref{geneva}}
        \and J.M.~Almenara\orcid{0000-0003-3208-9815}\inst{\ref{geneva},\ref{grenoble}}
        \and M.~Lendl\orcid{0000-0001-9699-1459}\inst{\ref{geneva}}
        \and D.~Ehrenreich\orcid{0000-0001-9704-5405}\inst{\ref{geneva}}
        \and F.~Bouchy\orcid{0000-0002-7613-393X}\inst{\ref{geneva}}
        \and X.~Bonfils\orcid{0000-0001-9003-8894}\inst{\ref{grenoble}}
        \and R.~Dancikova\inst{\ref{epfl}}
        \and A.~Deline\inst{\ref{geneva}}
        \and S.~Khan\orcid{0000-0001-5998-5885}\inst{\ref{epfl}}
        \and H.~Netzel\orcid{0000-0001-5608-0028}\inst{\ref{epfl},\ref{ncac}}
        \and M.~Shinde\orcid{0000-0002-8024-3779}\inst{\ref{geneva}}
        \and A.~Verdier\orcid{0009-0007-3803-2760}\inst{\ref{epfl}}
        }
   \institute{
        Geneva Observatory, University of Geneva, Chemin Pegasi 51, 1290 Versoix, Switzerland\label{geneva}
        \and Univ. Grenoble Alpes, CNRS, IPAG, F-38000 Grenoble, France\label{grenoble}
        \and Institute of Physics, \'Ecole Polytechnique F\'ed\'erale de Lausanne (EPFL), Observatoire de Sauverny, Chemin Pegasi 51b, 1290 Versoix, Switzerland\label{epfl}
        \and Nicolaus Copernicus Astronomical Centre, Polish Academy of Sciences, Bartycka 18, PL-00-716 Warszawa, Poland\label{ncac}
    }

   \date{}

 
  \abstract
   {
   TOI-3884\,b is a sub-Saturn transiting a fully convective M-dwarf. Observations indicate that the transit shape is chromatic and asymmetric as a result of persistent starspot crossings. This, along with the lack of photometric variability of the host star, indicates that the rotational axis of the star is tilted along our line of sight and the planet-occulted starspot is located close to the stellar pole. We acquired photometric transits over a period of three years with the Swiss 1.2-meter Euler telescope to track changes in the starspot configuration and detect any signs of decay or growth. The shape of the transit changes over time, and so far no two observations match perfectly. We conclude that the observed variability is likely not caused by changes in the temperature and size of the spot, but due to a slight (5.64 $\pm$ 0.64$^{\circ}$) misalignment between the spot center and the stellar pole, i.e., a small spin-spot angle ($\Theta$). In addition, we were able to obtain precise measurements of the sky-projected spin-orbit angle ($\lambda$) of 37.3 $\pm$ 1.5\degree, and the true spin-orbit angle ($\psi$) of 54.3 $\pm$ 1.4\degree. The precise alignment measurements along with future atmospheric characterisation with the James Webb Space Telescope will be vital for understanding the formation and evolution of close-in, massive planets around fully convective stars.}

   \keywords{stars: individual: \object{TOI-3884} --
            stars: low-mass --
            (stars:) starspots --
            (stars:) planetary systems --
            techniques: photometric
            }

   \maketitle
%

\section{Introduction}

With its nearly full-sky coverage, the Transiting Exoplanet Survey Satellite \citep[TESS,][]{Ricker2015} mission is surveying the sky for transiting planets, detecting even rare systems where unusual planetary and stellar properties lead to anomalous transit shapes. One of them is TOI-3884\,b \citep{Almenara2022,Libby-Roberts2023}, a 6.6-\REarth planet that transits an M4 dwarf star with a large polar spot. The main properties of the system are summarized in Table~\ref{table:adopted_params}. This spot-crossing configuration leads to transits that are highly asymmetric, showing a pronounced wavelength-dependent plateau during the first half of the transit as the planet passes across the (redder and less emissive) spot. 

Using TESS and ground-based observations, \citet{Almenara2022} found that the timing and amplitude of the spot crossings remained  consistent for at least two years, and subsequent follow-up observations reported slight shape variations between two transits \citep{Libby-Roberts2023}. The persistent spot crossings along with the lack of photometric variability of the host star suggest that the stellar spin axis is tilted along our line of sight, and the planet is occulting a spot located on or near the stellar rotational pole. Just before submission of this article, two independent articles \citet{Mori2025} and \citet{Tamburo2025} analysed photometric transits that confirmed shape variations at short timescales. 

TOI-3884 is a treasure trove for studying stellar activity, planetary atmospheres, and the observational connection between these phenomena. 
TOI-3884\,b is among the most favourable planets of its size for transit spectroscopy due to the large planet-to-star radius ratio and the host star’s relative brightness. Stellar activity is known to affect the observed transmission spectra \citep{Pont2008, Rackham2018, sage}, superposing a stellar contamination component on the planetary signal. In the case of highly-active M-dwarfs, this has hindered the identification of planetary atmospheric components \citep{Moran2023, Lim2023}.
With its well-characterised spot, TOI-3884 provides a unique opportunity to improve techniques for activity mitigation in transmission spectroscopy. This is the goal of two accepted JWST \citep{Gardner2006} programs in cycle three \citep{Garcia2024,Murray2024}, totalling seven transit observations of TOI-3884\,b.

The system also presents a unique target to study the properties and evolution of M-star stellar activity. For this purpose, techniques such as Doppler imaging and interferometry are commonly used \citep[e.g.,][]{Rottenbacher2016,Willamo2022}. However, both techniques have limitations. Doppler imaging is mainly applicable to rapidly rotating stars and is unable to determine the hemispheric location of the active regions, and interferometry is applicable only to large nearby stars and has a limited stellar surface resolution, making it difficult to track small-scale changes. Persistent spot-crossing events by transiting exoplanets, as observed for TOI-3884, can be used to study the evolution of active regions on the stellar surface through the photometric light curve anomalies \citep{Sanchis-Ojeda2013}. In addition, they can be used to determine the obliquity of the planetary orbit, providing insight into the system architectures \citep{Sanchis-Ojeda2011, Mocnik2016}. 

In this paper, we report new transit observations of TOI-3884\,b, tracking the changes in spot configuration over a three-year period to investigate short- or long-term temporal variations in the transit shape. The article is organized as follows: we present the new transit photometry of TOI-3884\,b in Sect.~\ref{section:observations}. We describe our data analysis and modelling in Sect.~\ref{section:analysis}, and report results in Sect.~\ref{section:results}. Finally, in Sect.~\ref{section:discussion} we discuss our findings, before concluding in Sect.~\ref{sec: conclusion}.

\begin{table}[h]
\small
  \setlength{\tabcolsep}{2pt}
      \caption[]{Adopted system parameters.}
         \label{table:adopted_params}
         \begin{tabular}{lcc}
            \hline
            \noalign{\smallskip}
            Parameter & \text{Value} & \text{Refs} \\
            \noalign{\smallskip}
            \hline
            \noalign{\smallskip}
            \textit{Stellar parameters} \\
            \noalign{\smallskip}
            Spectral type & M4 & 1 \\
            Stellar mass, $M_{\star}$ (\Msun) & $0.298 \pm 0.018$ & 1 \\
            Stellar radius, $R_{\star}$ (\Rsun) & $0.302 \pm 0.012$ & 1 \\
            Effective temperature, $T_{\rm eff}$ (K) & $ 3269 \pm 70$ & 2 \\
            Rotational period, $P_{\rm rot}$ (days) & $11.05 \pm 0.05$ & 3 \\
            \noalign{\medskip}
            
            \textit{TOI-3884\,b parameters} \\
            \noalign{\smallskip}
            Orbital period (days) & $ 4.54458373\pm0.00000067 $ & 4\\
            Mid-transit time (BJD$_{\rm TDB}$) & $ 2460351.818108\pm0.000070 $ & 4\\
            Orbital inclination (\degree) & 89.936$^{+0.047}_{-0.073}$ & 4 \\
            Scaled Semi-major axis (a/R$_{\star}$) & 25.7 $\pm$ 0.1 & 4\\
            Impact parameter, b & 0.03$^{+0.03}_{-0.02}$ & 4 \\
            Transit duration, T$_{\rm{14}}$ [hr] & 1.622 $\pm$ 0.006& 4 \\
            Planet-to-star radius ratio (R$_{p}$/R$_{\star}$) & 0.200 $\pm$ 0.004 & 4 \\
            Radius, $R_{\mathrm{p}}$ (\REarth) &  6.6 $\pm$ 0.4 & 4 \\
            Mass, $M_{\mathrm{p}}$ (\MEarth) & $32.6 \pm 7.4$ & 1 \\
            \hline
        \end{tabular}
        \begin{tablenotes}
        \small
        \item References : 1) \cite{Libby-Roberts2023}, 2) \citet{Almenara2022}, 3) \citet{Mori2025}, 4) This work.
        \end{tablenotes}
\end{table}

\section{Observations}\label{section:observations}

We observed six transits of TOI-3884\,b with the EulerCam \citep{Lendl2012} at the Swiss 1.2~m Euler telescope located at the La Silla Observatory, Chile. An observing log is presented in Table~\ref{table: observations}. The reduction consists in standard image correction including bias, flat and over-scan correction, followed by subsequent relative aperture photometry. The EulerCam reduction pipeline uses the \prose \citep{Garcia2022} framework that utilises \astropy \citep{astropy2013,astropy2018,astropy2022} and \photutils \citep{Bradley2023}. The optimal differential photometry is performed following \citet{Broeg2005}. Our observations were taken with the Next Generation Transit Survey (NGTS) broadband filter that has a bandpass from 520 to 890 nm \citep{Wheatley2018}. Thanks to its broad bandpass, this filter is optimal for observing faint stars such as TOI-3884 (G = 14.25). The observation on 2023-05-09 was affected by clouds at the end of the transit, and we removed all flux measurements beyond 2460074.62623 BJD$_{\rm{TDB}}$ from our analysis.

In addition, we also used three sectors of TESS observations: 22, 46 and 49, and ground-based observations from \citet{Almenara2022} for ephemeris refinement. The TESS light curves were downloaded from the Milkulski Archive for Space Telescopes (MAST), and corrected for instrumental systematics by the TESS Science Processing Operations Center \citep{Cladwell2020}. We used the Presearch Data Conditioning Simple Aperture Photometry (PDCSAP) light curves. The quality of the TESS light curves is not sufficient for constraining the transit shape of individual transits due to the faint target magnitude and the small size of the TESS telescope. The light curves obtained with the 1.2-meter Euler telescope are better suited to this purpose. The root-mean-square (RMS) scatter of our light curves, calculated over five minutes, is reported in Table \ref{table: observations}. For comparison, the RMS scatter of TESS light curves is 3404 and 3370 ppm for Sectors 46 and 49, respectively.

\begin{table}[h]
\small
  \setlength{\tabcolsep}{5pt}
      \caption[]{Observing log of EulerCam observations of TOI-3884\,b}
         \label{table: observations}
         \begin{tabular}{ccccc}
            \hline
            \noalign{\smallskip}
            Date & Airmass & Average       & Exposure & RMS$_{\rm{5min}}$ \\
            {[UT]} & range   & FWHM {[pixel]} & time {[second]} & [ppm]\\
            \noalign{\smallskip}
            \hline
            \noalign{\smallskip}
            2023-05-09 & 1.3 - 1.4 & 4.55 & 60 - 120 & 1477 \\
            2024-02-10 & 1.3 - 1.5 & 3.81 & 30 - 40 & 1731\\
            2024-03-13 & 1.3 - 2.1 & 4.93 & 30 - 60 & 1786\\
            2025-02-03 & 1.3 - 1.5 & 6.29 & 68 - 90 & 1457\\
            2025-03-07 & 1.3 - 2.2 & 5.77 & 60 - 120 & 1585\\
            2025-03-16 & 1.3 - 2.0 & 4.36 & 40 - 50 & 1757\\
            \hline
            \small
        \end{tabular}
\end{table}
 
\begin{figure*}
   \centering
\resizebox{\textwidth}{!}{\includegraphics[trim=0.0cm 0.0cm 0.0cm 0.0cm]{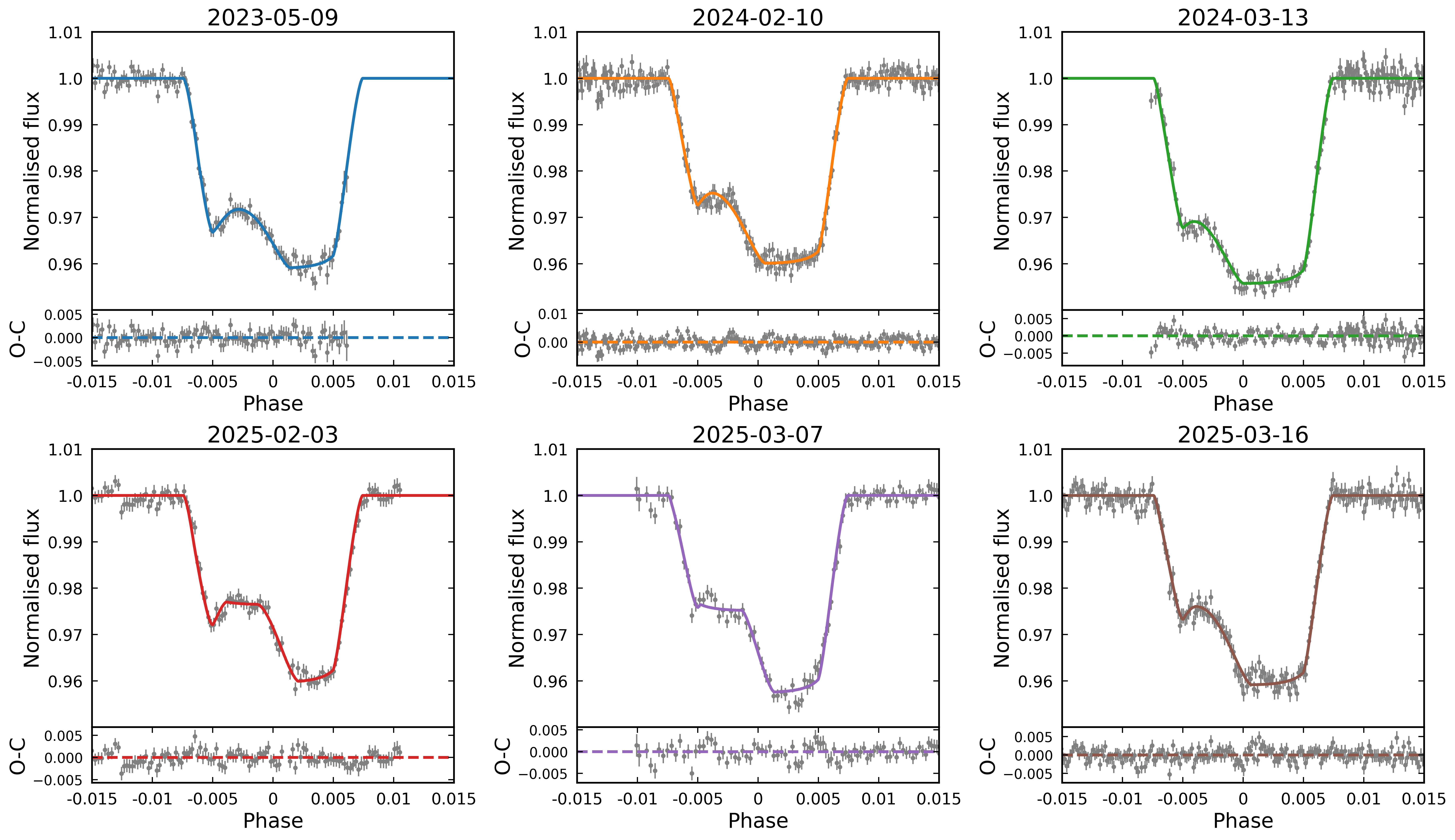}}
      \caption{Phase-folded EulerCam light curves of TOI-3884\,b. The top panel shows the normalised flux, corrected for instrumental systematics, with the best-fit \pytranspot models, obtained using the second (or the spin-spot misalignment) approach over-laid on top. The lower panels displays the model residuals.}
         \label{fig:lightcurves_approach_2}
\end{figure*}

\section{Transit and spot modelling}\label{section:analysis}

All EulerCam transits, shown in Figure \ref{fig:lightcurves_approach_2}, are asymmetric due to stellar spot crossing events. The data were collected over a three-year period and reveal variability in the shape of the transit. So far, no two transit observations match perfectly, which points to both short- and long-term variations in the properties of the polar spot. We aim to constrain these variations by modelling the overall changes in the transit shape. 

To refine the transit ephemeris, we updated the transit modelling using \juliet \citep{Espinoza2019,Speagle2020}, as presented in Sect.~4 of \citet{Almenara2022}, incorporating the new EulerCam observations and six new transits observed with ExTrA \citep{Bonfils2015} between 2023 and 2025. The new ExTrA data was acquired and reduced with the same setup as in  \cite{Almenara2022}. We used the 8 arcsecond aperture fibres, the low-resolution mode of the spectrograph (R $\sim$ 20), and 60 second exposure time. In addition to the transit model \citep{Kreidberg2015}, we applied Gaussian process (GP) regression model with an approximate Matern kernel \citep[implemented in \texttt{celerite},][]{Foreman-mackey2017} to account for the spot crossing. We choose log-uniform priors on the GP hyperparameters: normalized flux amplitude [$10^{-6}$, 0.1] and time-scale [$10^{-3}$, 10] days. The ephemeris obtained is presented in Table~\ref{table:adopted_params} and the light curves are shown in Fig. \ref{fig:juliet}. We obtain a $1\sigma$ timing uncertainty between 5 and 10~seconds for our EulerCam observations and a period uncertainty of 73 milliseconds. Based on this, we fixed the mid-transit time and period to the derived values for the remainder of our analysis. Additionally, we kept the eccentricity fixed to zero, as our data do not have the extremely high precision necessary to detect eccentricity from transit light curves.

For spot modelling, we analyse the transit light curves with \texttt{PyTranSpot} \citep{Juvan2018,sage}. The tool uses a pixelation approach to project a stellar sphere, a transit chord, and stellar spots or faculae on a two-dimensional Cartesian grid. The active regions are assumed to be circular and of uniform brightness. The model light curve is parametrised using the mid-transit time, planet-to-star radius ratio, orbital inclination, scaled semimajor axis, orbital period, orbital eccentricity, argument of periastron and quadratic limb-darkening coefficients. The spot effect is parametrised using their position (co-latitude and longitude), size, and contrast with respect to the unspotted stellar photosphere. The intersection of the line of sight with the stellar surface corresponds to a co-latitude of 90\degree\ and a longitude of 0\degree, with the normal to the planet's orbital plane pointing toward a co-latitude of 0\degree. We used the \texttt{emcee} algorithm \citep{Goodman2010, emcee} to sample the posterior distribution. We modelled the transit light curves using two different spot-modelling approaches. 

The first approach involves joint modelling of individual transit light curves, allowing the spot parameters to vary freely for each observation. This approach enables the measurement of temporal variations in the properties of the spot to track its decay or growth \citep{Namekata2019,Namekata2020}. We set broad uniform priors on the planetary parameters including semimajor axis, inclination, and limb-darkening coefficients. Additionally, for each observation, we set broad uniform priors on longitude and co-latitude to cover the entire stellar surface. The angular size and contrast of the spot vary between 0\degree\ and 90\degree\ and between zero and one, respectively. A spot contrast of one corresponds to a spot with the same temperature as the unspotted photosphere.

The second approach involves assuming static spot properties, i.e, the same spot size and contrast over time. We jointly fit for a global spot size and contrast, while allowing the spot locations on the projected disk to vary for individual transit observations. This scenario is motivated by \citet{Giles2017}, who demonstrated using \emph{Kepler} light curves that large starspots exhibit slow decay rates regardless of the stellar spectral type. Therefore, under the assumption that the starspot is not evolving, the observed variability may be attributed to the varying position of the spot on the stellar surface resulting from stellar rotation. We adopted the same set of broad uniform priors as used in the previous approach. The best-fit model for each observation using this approach is shown in Fig.~\ref{fig:lightcurves_approach_2}, and using the first approach is shown in Fig.~\ref{fig:lightcurves_approach_1}.

To account for any instrumental systematics, we always simultaneously fitted a photometric baseline model to each light curve. In all cases, the baseline consisted in two second-order polynomials, one in airmass and second one in stellar FWHM (full width at half-maximum) or in exposure time. The choice of the baseline was made by minimising the root-mean-square of the residuals. In addition, we include a dilution term to account for the observed variability in transit depths, likely resulting from temporal changes in the unocculted active regions \citep{Szabo2021,Wang2024}. Lastly, a multiplicative jitter term is used to scale the flux uncertainties. 

\section{Results}\label{section:results}

\begin{figure}[h]
  \centering
  \includegraphics[width=0.49\textwidth]{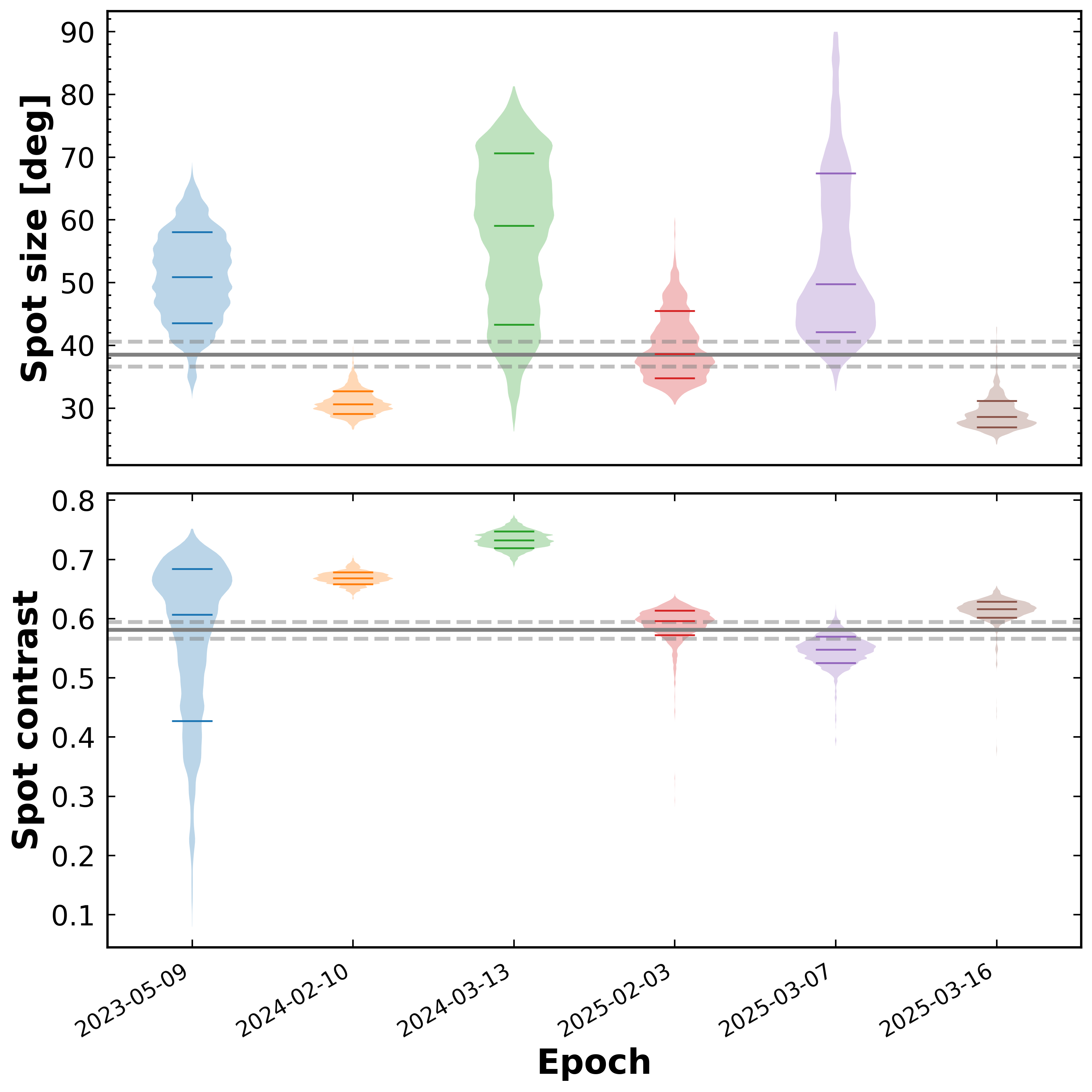}
  \caption{Posteriors for spot size and contrast for TOI-3884. The top panel shows a violin plot with spot size and posterior distribution for each observation obtained using the first (or spot evolution) approach. The lower plot displays the spot contrast. The horizontal lines show the median spot size and contrast obtained using the second (or the spin-spot misalignment) approach, with dashed lines showing the 1$\sigma$ limits.} \label{figure:spot_size_temp}
\end{figure}

\subsection{Spot evolution}\label{subsection:evolution}

Active regions on the stellar surface are known to evolve over time, with smaller spots decaying more rapidly than larger ones \citep{Giles2017}. The decay rate depends on multiple factors, including stellar rotation, differential rotation, and spectral type \citep{Strassmeier2009,Basri2022}.  
Polar cap-like spots have a different formation history than low- to mid-latitude spots, and thus their lifetimes are not comparable \citep{Strassmeier2009}. The lifetimes of large spots are limited by the strength or shear of differential rotation, which ultimately leads to their fragmentation into smaller spots. Observations suggest that the strength of differential rotation decreases steadily with decreasing stellar surface temperature \citep{kuker2008}. For fully convective stars like TOI-3884, the long convective-turnover timescales decreases the strength of differential rotation \citep{Kitchatinov2011}. Thus, a differential rotation-induced decay of the polar spot of TOI-3884 is unlikely. 

Our analysis using the first approach measures the spot properties for each observation individually, to track their evolution over time without any prior assumptions on their stability over time. As shown in Fig.~\ref{figure:spot_size_temp}, we do not observe steady decay or growth in the measured size or contrast of the spot. Assuming that both the stellar spot and the photosphere are emitting as blackbodies, we used Equation~1 of \citet{Silva2003} to convert the derived spot contrast into spot temperature. For this, we assumed that the effective temperature of the star (T$_{\rm{eff}}$ = 3269~K, \citealp{Almenara2022}) is a suitable proxy for the temperature of the clear photosphere. The derived spot temperatures range from 2700~K to 3000~K, and the spot sizes (i.e., the angular diameter) vary from 28\degree\ to 59\degree. Our finding of a globally stable spot size and contrast supports the view of globally stable polar spots on the fully-convective M-dwarf TOI-3884. As expected, spot size and contrast are strongly correlated, leading to significant uncertainty when measuring them from individual observations. A more precise contrast measurement can be achieved by stacking multiple observations, ideally in different filters, to break the size-contrast degeneracy.

\subsection{Misalignment between stellar spin axis and polar spot}\label{subsection:obliquity}

\begin{figure}[!ht]
  \centering
  \includegraphics[width=0.49\textwidth]{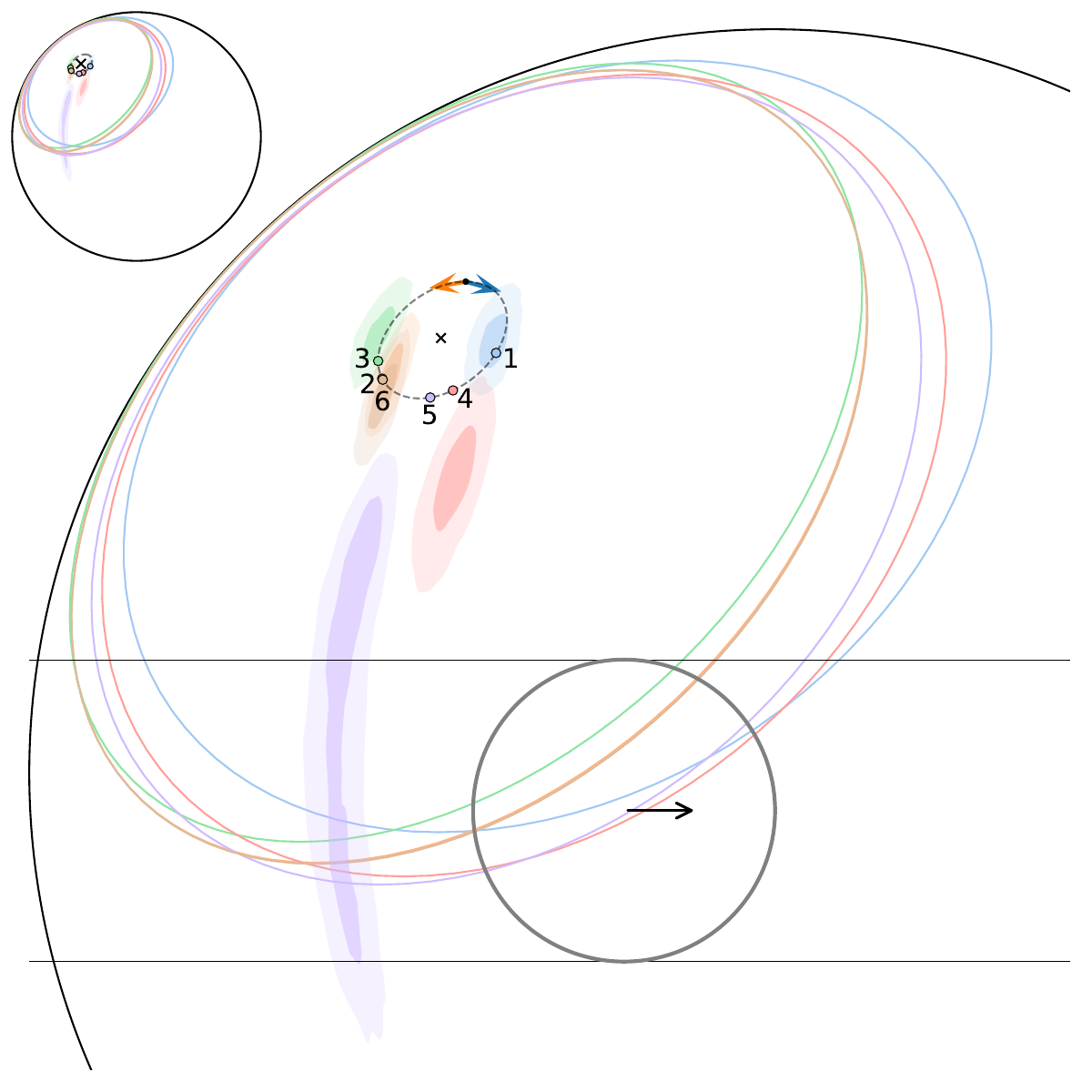}
  \caption{The posterior position of the spot centre on the stellar surface, as inferred from the second approach analysis for each transit observation, is shown with shaded regions indicating the $1\sigma$ and $2\sigma$ credible intervals, using the same colour coding as in Figs.~\ref{fig:lightcurves_approach_2} and  \ref{figure:spot_size_temp}. The transit path for the best-fit model is represented by horizontal lines, while the planet, depicted as a circle, contains an arrow indicating its direction of movement. The best-fit spherical circle for the spin-spot misalignment model is shown with a dashed gray line, with coloured dots representing the positions of the transit observations. The coloured circles centred on these points represent the spot. The black dot in the spherical circle indicates the position at the reference time, while the black diagonal cross marks the position of the spin axis. An orange arrow represents CCW movement, while a blue arrow represents CW movement.} \label{figure:magnetic_obliquity}
\end{figure}

\begin{table}[!ht]
    \tiny
    \renewcommand{\arraystretch}{1.25}
    \setlength{\tabcolsep}{1pt}
\centering
\caption{Spin-spot misalignment model for CCW and P$_{\rm rot}\sim$10.8~days.}\label{table:magnetic_obliquity}
\begin{tabular}{lccc}
\hline
Parameter & Units & Prior & Posterior median   \\
&  & & and $1\sigma$ \\
\hline
Center latitude & [\degree] & $U(-90, 90)$ & $35.7 \pm 1.4$  \\
Center longitude & [\degree] & $U(-180, 180)$ & $-33.3 \pm 1.1$  \\
Radius (spin-spot angle, $\Theta$) & [\degree] & $U(0, 90)$ & $5.64 \pm 0.64$  \\
Rotational period (P$_{\rm rot}$) & [days] & $U(10, 12)$ & $10.8072 \pm 0.0098$$^\dagger$  \\
Reference time$^\ddagger$\;-\;2\;460\;000 & [BJD$_{\mathrm{TDB}}$] & $U(63.8, 85.4)$ & $66.90 \pm 0.66$$^\dagger$  \\
Inclination of the spin axis, $i_\star$ & [\degree] &  & $47.3 \pm 1.2$ \\
Sky-projected spin-orbit angle, $\lambda$ & [\degree] &  & $37.3 \pm 1.5$ \\
True spin–orbit angle, $\psi$ & [\degree] & & $54.3 \pm 1.4$ \\
\hline
\end{tabular}
\tablefoot{$U$(l,u): Uniform distribution prior in the range [l, u]. $^\dagger$ Multimodal, only the higher CCW posterior peak is shown. $^\ddagger$ The position of the spot at the reference time is marked with a black dot in Fig.~\ref{figure:magnetic_obliquity}. Sky-projected and true spin-orbit angles are calculated based on the spin axis orientation and the orbital inclination.}
\end{table}

While temporally stable, the spot centre might still be slightly offset from the stellar rotational pole. The rotation of the spot around the stellar spin axis, and thus a slight offset in its projected position during transit, then produces the observed changes in the transit light curves. Using the second analysis approach described in Section~\ref{section:analysis}, i.e., assuming constant spot size and contrast, we estimated the spot location for each transit observation. The posteriors are presented in Table~\ref{table:spot_parameters_approach2} and in Fig.~\ref{figure:magnetic_obliquity}. The angle between the centre of the spot and the stellar rotation axis represents the spin-spot angle. To derive this quantity, we modelled these time series of spot centre latitude and longitude (see Fig.~\ref{figure:model}) using a spherical circle defined by its radius (i.e., the spin-spot angle), the latitude and longitude of its centre, rotational period, and a reference time for a given phase. We used uniform priors (see Table~\ref{table:magnetic_obliquity}) and the \texttt{emcee} algorithm \citep{Goodman2010, emcee} to sample the posterior distribution. 

For the rotation period, we choose uniform priors between 10 and 12 days. This is motivated by a reanalysis of TESS light curves where we find evidence of long-term flux variability with a period of $\sim$10 days. TOI-3884 was observed by TESS in three different sectors: 22, 46, and 49. We downloaded the TESS light curves using the \texttt{lightkurve} package \citep{lightcurve2018}. Sector 22 was excluded from our analysis, as the light curves generated by the TESS-SPOC pipeline are not available for this sector. For sectors 46 and 49, we removed all flux measurements with a non-zero quality flag, and used the Pre-search Data Conditioning Simple Aperture Photometry (PDC-SAP) light curves. We searched for any prominent periodicities in the processed light curves using the Lomb-Scargle algorithm \citep{lomb, scargle}. In sector 46, we did not detect any significant periodicities, likely due to high systematic noise. However, in sector 49, we measured a peak periodicity corresponding to a rotation period of 10.2 days (Fig.~\ref{fig:tess_sector_49}). In addition, our priors cover the rotation period of 11.043 days reported by \citet{Mori2025}. 

We explored both directions of motion along the spherical circle, clockwise (CW) and counter-clockwise (CCW), see Fig.~\ref{figure:magnetic_obliquity}. The posterior for the rotational period presents several peaks (see Fig.~\ref{figure:Prot}), but all have compatible posteriors for the latitude and longitude of the spherical circle's centre and its radius, the spin-spot angle, for which we obtained a value of $5.64 \pm 0.64$\degree\  (Table~\ref{table:magnetic_obliquity}). The best model for a rotational period of $\sim10.8$~days and CCW direction is plotted in Fig.~\ref{figure:magnetic_obliquity}. 

We performed a model comparison by calculating the Bayesian Information Criterion (BIC) for two different approaches. The BIC accounts for the number of free parameters, with a lower BIC indicating a more preferred model. The BIC for the spot evolution approach is 1571 and for the spin-spot misalignment approach is 1522. This highlights that the latter approach is preferred. In addition, the spot locations inferred from the spot evolution approach are 2.5$\sigma$ consistent with the locations inferred by the spin-spot misalignment model for all observations. Thus, our measurement of the spin-spot angle is not biased by fitting a static spot size and contrast for all observations.

The measured misalignment between the starspot position and the stellar pole is expected to produce out-of-transit flux variability due to stellar rotation. We used the \texttt{SAGE} tool \citep{sage} to generate predicted light curves, assuming a rotation period of $\sim$10.8 days. The modelled flux variability has an amplitude of approximately 2\%, which is significantly higher than the observed flux variability in different TESS sectors. However, the amplitude is compatible with recent and more precise ground-based photometric monitoring \citep{Mori2025}.

\section{Discussion}\label{section:discussion}
We find that, while the projected position of the starspot displays small changes, its location near the rotational pole, size of $\sim$40$^{\circ}$ and contrast have remained largely stable since its first observation by TESS in 2020 \citep{Almenara2022}. This stability of a polar spot on a fully convective M-dwarf is in line with  photometric studies of M-star variability \citep{Giles2017} and also observations by \citet{Davenport2015}, who reported starspots persisting for several years in an individual object.

Spectropolarimetric surveys utilising the Zeeman-Doppler imaging technique have mapped the surface brightness and magnetic topologies of multiple M dwarfs (e.g. \citealp{Morin2008, Donati2008, Morin2010, Kochukhov2017, Klein2021, Bellotti2023, hebrard2016, See2025}). For fully convective stars similar to TOI-3884 (M$_{\star}\lesssim$0.34 M$_{\odot}$), studies have found both large-scale dipolar fields and weak non-axisymmetric fields \citep{Morin2010}. This apparent dichotomy may not be due to intrinsic differences between different stars, as multi-season observations of several objects have shown that magnetic field configurations can evolve from strong-dipolar to weak, non-axisymmetric or toroidal and vice versa (e.g. \citealp{Donati2023, Bellotti2024}). For several M dwarfs with poloidal fields, magnetic obliquities, i.e. the misalignment angle between the stellar rotation and magnetic axes, have been measured. For Proxima Cen, a large obliquity of 51$^{\circ}$ was found \citep{Klein2021}, whereas for GJ436  and AU Mic, smaller misalignments of 15$^{\circ}$ and 20$^{\circ}$ were found \citep{Bellotti2023,Donati2025}. It is not unlikely that TOI-3884's large spot closely traces the location of the stellar magnetic pole. In this case, this would indicate a dipolar field, with a nearly aligned geometry, consistent with other objects near the upper boundary of the fully convective stellar mass range, such as GJ358 \citep{hebrard2016}.

The misaligned orbit of the planet can lead to excess heating via magnetic induction \citep{Kislyakova2017}. For fully convective M dwarfs, the magnetic field strength can be on the order of kilogauss \citep{Vidotto2014}. For TOI-3884 b, the normal of the orbital plane is inclined with respect to the stellar spin axis. This configuration can result in a variable magnetic field due to orbital motion, leading to strong heating of the planet \citep{Kislyakova2017, Kislyakova2018}. 

To further constrain the spin-spot angle, we require transit observations targeting different phases of spin-spot misalignment. There are limited observing windows from the ground, but high-precision space-based observations from CHEOPS \citep{benz_cheops} could be vital to refining the measurement of the spin-spot angle.

\section{Conclusion}\label{sec: conclusion}

We followed transits of TOI-3884\,b over a period of three years to track changes in the planet-occulted starspot located close to the stellar pole. Our key findings are summarised below:

\begin{enumerate}
    \item We observed variability in the shape, i.e. timing and amplitude of the planet-occulted starspot in the transit light curves of TOI-3884\,b.
    
    \item We find that the observed shape variability is best explained by a starspot with stable temperature and size that is slightly offset from the stellar rotation pole. Its precise position during transit defines the shape of the light curve. We measured the starspot position offset angle relative to the rotation axis (i.e., the spin-spot angle), and found it to be 5.64 $\pm$ 0.64\degree. Assuming the starspot is located at or near the stellar magnetic pole, this would point towards a poloidal field similar to that observed for other fully-convective M-stars \citep{Donati_2010}.

    \item The recurring starspot crossing allow us to constrain the planetary architecture by measuring the sky-projected spin-orbit angle of 37.3 $\pm$ 1.5\degree\ (or 142.7 $\pm$ 1.5\degree), and the true spin-orbit angle of 54.3 $\pm$ 1.4\degree\ (or 125.7 $\pm$ 1.4\degree). This points to TOI-3884\,b having undergone a migration process capable of producing significant orbital misalignment of the planet. Alternatively, this misalignment could also result from the host star tilting its spin axis relative to the position of the protoplanetary disk \citep{Batygin2012}.
\end{enumerate}


\begin{acknowledgements}
HC and ML acknowledge support of the Swiss National Science Foundation under grant number PCEFP2\_194576. This work has been carried out within the framework of the NCCR PlanetS supported by the Swiss National Science Foundation under grants 51NF40\_182901 and 51NF40\_205606. This paper includes data collected by the TESS mission. Funding for the TESS mission is provided by the NASA’s Science Mission Directorate. We acknowledge funding from the European Research Council under the ERC Grant Agreement n. 337591-ExTrA. DEH and ADE acknowledge the financial support of the National Centre of Competence in Research PlanetS supported by the Swiss National Science Foundation under grants 51NF40\_182901 and 51NF40\_205606. HN acknowledges support from the European Research Council (ERC) under the European Union’s Horizon 2020 research and innovation program (grant agreement No. 951549 - UniverScale).
\end{acknowledgements}

%
%

\bibliographystyle{aa}
\bibliography{TOI-3884_ECAM}

\begin{thebibliography}{66}
\expandafter\ifx\csname natexlab\endcsname\relax\def\natexlab#1{#1}\fi

\bibitem[{{Almenara} {et~al.}(2022){Almenara}, {Bonfils}, {Forveille}, {Astudillo-Defru}, {Ciardi}, {Schwarz}, {Collins}, {Cointepas}, {Lund}, {Bouchy}, {Charbonneau}, {D{\'\i}az}, {Delfosse}, {Kidwell}, {Kunimoto}, {Latham}, {Lissauer}, {Murgas}, {Ricker}, {Seager}, {Vezie}, \& {Watanabe}}]{Almenara2022}
{Almenara}, J.~M., {Bonfils}, X., {Forveille}, T., {et~al.} 2022, \aap, 667, L11

\bibitem[{{Astropy Collaboration} {et~al.}(2022){Astropy Collaboration}, {Price-Whelan}, {Lim}, {Earl}, {Starkman}, {Bradley}, {Shupe}, {Patil}, {Corrales}, {Brasseur}, {N{"o}the}, {Donath}, {Tollerud}, {Morris}, {Ginsburg}, {Vaher}, {Weaver}, {Tocknell}, {Jamieson}, {van Kerkwijk}, {Robitaille}, {Merry}, {Bachetti}, {G{"u}nther}, {Aldcroft}, {Alvarado-Montes}, {Archibald}, {B{'o}di}, {Bapat}, {Barentsen}, {Baz{'a}n}, {Biswas}, {Boquien}, {Burke}, {Cara}, {Cara}, {Conroy}, {Conseil}, {Craig}, {Cross}, {Cruz}, {D'Eugenio}, {Dencheva}, {Devillepoix}, {Dietrich}, {Eigenbrot}, {Erben}, {Ferreira}, {Foreman-Mackey}, {Fox}, {Freij}, {Garg}, {Geda}, {Glattly}, {Gondhalekar}, {Gordon}, {Grant}, {Greenfield}, {Groener}, {Guest}, {Gurovich}, {Handberg}, {Hart}, {Hatfield-Dodds}, {Homeier}, {Hosseinzadeh}, {Jenness}, {Jones}, {Joseph}, {Kalmbach}, {Karamehmetoglu}, {Ka{l}uszy{'n}ski}, {Kelley}, {Kern}, {Kerzendorf}, {Koch}, {Kulumani}, {Lee}, {Ly}, {Ma}, {MacBride}, {Maljaars}, {Muna}, {Murphy}, {Norman}, {O'Steen},
  {Oman}, {Pacifici}, {Pascual}, {Pascual-Granado}, {Patil}, {Perren}, {Pickering}, {Rastogi}, {Roulston}, {Ryan}, {Rykoff}, {Sabater}, {Sakurikar}, {Salgado}, {Sanghi}, {Saunders}, {Savchenko}, {Schwardt}, {Seifert-Eckert}, {Shih}, {Jain}, {Shukla}, {Sick}, {Simpson}, {Singanamalla}, {Singer}, {Singhal}, {Sinha}, {Sip{H{o}}cz}, {Spitler}, {Stansby}, {Streicher}, {{{S}}umak}, {Swinbank}, {Taranu}, {Tewary}, {Tremblay}, {Val-Borro}, {Van Kooten}, {Vasovi{'c}}, {Verma}, {de Miranda Cardoso}, {Williams}, {Wilson}, {Winkel}, {Wood-Vasey}, {Xue}, {Yoachim}, {Zhang}, {Zonca}, \& {Astropy Project Contributors}}]{astropy2022}
{Astropy Collaboration}, {Price-Whelan}, A.~M., {Lim}, P.~L., {et~al.} 2022, \apj, 935, 167

\bibitem[{{Astropy Collaboration} {et~al.}(2018){Astropy Collaboration}, {Price-Whelan}, {Sip{\H{o}}cz}, {G{\"u}nther}, {Lim}, {Crawford}, {Conseil}, {Shupe}, {Craig}, {Dencheva}, {Ginsburg}, {Vand erPlas}, {Bradley}, {P{\'e}rez-Su{\'a}rez}, {de Val-Borro}, {Aldcroft}, {Cruz}, {Robitaille}, {Tollerud}, {Ardelean}, {Babej}, {Bach}, {Bachetti}, {Bakanov}, {Bamford}, {Barentsen}, {Barmby}, {Baumbach}, {Berry}, {Biscani}, {Boquien}, {Bostroem}, {Bouma}, {Brammer}, {Bray}, {Breytenbach}, {Buddelmeijer}, {Burke}, {Calderone}, {Cano Rodr{\'\i}guez}, {Cara}, {Cardoso}, {Cheedella}, {Copin}, {Corrales}, {Crichton}, {D'Avella}, {Deil}, {Depagne}, {Dietrich}, {Donath}, {Droettboom}, {Earl}, {Erben}, {Fabbro}, {Ferreira}, {Finethy}, {Fox}, {Garrison}, {Gibbons}, {Goldstein}, {Gommers}, {Greco}, {Greenfield}, {Groener}, {Grollier}, {Hagen}, {Hirst}, {Homeier}, {Horton}, {Hosseinzadeh}, {Hu}, {Hunkeler}, {Ivezi{\'c}}, {Jain}, {Jenness}, {Kanarek}, {Kendrew}, {Kern}, {Kerzendorf}, {Khvalko}, {King}, {Kirkby}, {Kulkarni},
  {Kumar}, {Lee}, {Lenz}, {Littlefair}, {Ma}, {Macleod}, {Mastropietro}, {McCully}, {Montagnac}, {Morris}, {Mueller}, {Mumford}, {Muna}, {Murphy}, {Nelson}, {Nguyen}, {Ninan}, {N{\"o}the}, {Ogaz}, {Oh}, {Parejko}, {Parley}, {Pascual}, {Patil}, {Patil}, {Plunkett}, {Prochaska}, {Rastogi}, {Reddy Janga}, {Sabater}, {Sakurikar}, {Seifert}, {Sherbert}, {Sherwood-Taylor}, {Shih}, {Sick}, {Silbiger}, {Singanamalla}, {Singer}, {Sladen}, {Sooley}, {Sornarajah}, {Streicher}, {Teuben}, {Thomas}, {Tremblay}, {Turner}, {Terr{\'o}n}, {van Kerkwijk}, {de la Vega}, {Watkins}, {Weaver}, {Whitmore}, {Woillez}, {Zabalza}, \& {Astropy Contributors}}]{astropy2018}
{Astropy Collaboration}, {Price-Whelan}, A.~M., {Sip{\H{o}}cz}, B.~M., {et~al.} 2018, \aj, 156, 123

\bibitem[{{Astropy Collaboration} {et~al.}(2013){Astropy Collaboration}, {Robitaille}, {Tollerud}, {Greenfield}, {Droettboom}, {Bray}, {Aldcroft}, {Davis}, {Ginsburg}, {Price-Whelan}, {Kerzendorf}, {Conley}, {Crighton}, {Barbary}, {Muna}, {Ferguson}, {Grollier}, {Parikh}, {Nair}, {Unther}, {Deil}, {Woillez}, {Conseil}, {Kramer}, {Turner}, {Singer}, {Fox}, {Weaver}, {Zabalza}, {Edwards}, {Azalee Bostroem}, {Burke}, {Casey}, {Crawford}, {Dencheva}, {Ely}, {Jenness}, {Labrie}, {Lim}, {Pierfederici}, {Pontzen}, {Ptak}, {Refsdal}, {Servillat}, \& {Streicher}}]{astropy2013}
{Astropy Collaboration}, {Robitaille}, T.~P., {Tollerud}, E.~J., {et~al.} 2013, \aap, 558, A33

\bibitem[{{Basri} {et~al.}(2022){Basri}, {Streichenberger}, {McWard}, {Edmond}, {Tan}, {Lee}, \& {Melton}}]{Basri2022}
{Basri}, G., {Streichenberger}, T., {McWard}, C., {et~al.} 2022, \apj, 924, 31

\bibitem[{{Batygin}(2012)}]{Batygin2012}
{Batygin}, K. 2012, \nat, 491, 418

\bibitem[{{Bellotti} {et~al.}(2023){Bellotti}, {Fares}, {Vidotto}, {Morin}, {Petit}, {Hussain}, {Bourrier}, {Donati}, {Moutou}, \& {H{\'e}brard}}]{Bellotti2023}
{Bellotti}, S., {Fares}, R., {Vidotto}, A.~A., {et~al.} 2023, \aap, 676, A139

\bibitem[{{Bellotti} {et~al.}(2024){Bellotti}, {Morin}, {Lehmann}, {Petit}, {Hussain}, {Donati}, {Folsom}, {Carmona}, {Martioli}, {Klein}, {Fouqu{\'e}}, {Moutou}, {Alencar}, {Artigau}, {Boisse}, {Bouchy}, {Bouvier}, {Cook}, {Delfosse}, {Doyon}, \& {H{\'e}brard}}]{Bellotti2024}
{Bellotti}, S., {Morin}, J., {Lehmann}, L.~T., {et~al.} 2024, \aap, 686, A66

\bibitem[{{Benz} {et~al.}(2021){Benz}, {Broeg}, {Fortier}, {Rando}, {Beck}, {Beck}, {Queloz}, {Ehrenreich}, {Maxted}, {Isaak}, {Billot}, {Alibert}, {Alonso}, {Ant{\'o}nio}, {Asquier}, {Bandy}, {B{\'a}rczy}, {Barrado}, {Barros}, {Baumjohann}, {Bekkelien}, {Bergomi}, {Biondi}, {Bonfils}, {Borsato}, {Brandeker}, {Busch}, {Cabrera}, {Cessa}, {Charnoz}, {Chazelas}, {Collier Cameron}, {Corral Van Damme}, {Cortes}, {Davies}, {Deleuil}, {Deline}, {Delrez}, {Demangeon}, {Demory}, {Erikson}, {Farinato}, {Fossati}, {Fridlund}, {Futyan}, {Gandolfi}, {Garcia Munoz}, {Gillon}, {Guterman}, {Gutierrez}, {Hasiba}, {Heng}, {Hernandez}, {Hoyer}, {Kiss}, {Kovacs}, {Kuntzer}, {Laskar}, {Lecavelier des Etangs}, {Lendl}, {L{\'o}pez}, {Lora}, {Lovis}, {L{\"u}ftinger}, {Magrin}, {Malvasio}, {Marafatto}, {Michaelis}, {de Miguel}, {Modrego}, {Munari}, {Nascimbeni}, {Olofsson}, {Ottacher}, {Ottensamer}, {Pagano}, {Palacios}, {Pall{\'e}}, {Peter}, {Piazza}, {Piotto}, {Pizarro}, {Pollaco}, {Ragazzoni}, {Ratti}, {Rauer}, {Ribas}, {Rieder},
  {Rohlfs}, {Safa}, {Salatti}, {Santos}, {Scandariato}, {S{\'e}gransan}, {Simon}, {Smith}, {Sordet}, {Sousa}, {Steller}, {Szab{\'o}}, {Szoke}, {Thomas}, {Tschentscher}, {Udry}, {Van Grootel}, {Viotto}, {Walter}, {Walton}, {Wildi}, \& {Wolter}}]{benz_cheops}
{Benz}, W., {Broeg}, C., {Fortier}, A., {et~al.} 2021, Experimental Astronomy, 51, 109

\bibitem[{{Bonfils} {et~al.}(2015){Bonfils}, {Almenara}, {Jocou}, {Wunsche}, {Kern}, {Delboulb{\'e}}, {Delfosse}, {Feautrier}, {Forveille}, {Gluck}, {Lafrasse}, {Magnard}, {Maurel}, {Moulin}, {Murgas}, {Rabou}, {Rochat}, {Roux}, \& {Stadler}}]{Bonfils2015}
{Bonfils}, X., {Almenara}, J.~M., {Jocou}, L., {et~al.} 2015, in Society of Photo-Optical Instrumentation Engineers (SPIE) Conference Series, Vol. 9605, Techniques and Instrumentation for Detection of Exoplanets VII, 96051L

\bibitem[{Bradley(2023)}]{Bradley2023}
Bradley, L. 2023, astropy/photutils: 1.8.0

\bibitem[{{Broeg} {et~al.}(2005){Broeg}, {Fern{\'a}ndez}, \& {Neuh{\"a}user}}]{Broeg2005}
{Broeg}, C., {Fern{\'a}ndez}, M., \& {Neuh{\"a}user}, R. 2005, Astronomische Nachrichten, 326, 134

\bibitem[{{Caldwell} {et~al.}(2020){Caldwell}, {Tenenbaum}, {Twicken}, {Jenkins}, {Ting}, {Smith}, {Hedges}, {Fausnaugh}, {Rose}, \& {Burke}}]{Cladwell2020}
{Caldwell}, D.~A., {Tenenbaum}, P., {Twicken}, J.~D., {et~al.} 2020, Research Notes of the American Astronomical Society, 4, 201

\bibitem[{{Chakraborty} {et~al.}(2024){Chakraborty}, {Lendl}, {Akinsanmi}, {Petit dit de la Roche}, \& {Deline}}]{sage}
{Chakraborty}, H., {Lendl}, M., {Akinsanmi}, B., {Petit dit de la Roche}, D.~J.~M., \& {Deline}, A. 2024, \aap, 685, A173

\bibitem[{{Davenport} {et~al.}(2015){Davenport}, {Hebb}, \& {Hawley}}]{Davenport2015}
{Davenport}, J. R.~A., {Hebb}, L., \& {Hawley}, S.~L. 2015, \apj, 806, 212

\bibitem[{Donati(2010)}]{Donati_2010}
Donati, J.-F. 2010, Proceedings of the International Astronomical Union, 6, 23–31

\bibitem[{{Donati} {et~al.}(2025){Donati}, {Cristofari}, {Moutou}, {L'Heureux}, {Cook}, {Artigau}, {Alencar}, {Gaidos}, {Vidotto}, {Petit}, {Carmona}, \& {Ray}}]{Donati2025}
{Donati}, J.~F., {Cristofari}, P.~I., {Moutou}, C., {et~al.} 2025, \aap, 700, A227

\bibitem[{{Donati} {et~al.}(2023){Donati}, {Lehmann}, {Cristofari}, {Fouqu{\'e}}, {Moutou}, {Charpentier}, {Ould-Elhkim}, {Carmona}, {Delfosse}, {Artigau}, {Alencar}, {Cadieux}, {Arnold}, {Petit}, {Morin}, {Forveille}, {Cloutier}, {Doyon}, {H{\'e}brard}, \& {SLS Collaboration}}]{Donati2023}
{Donati}, J.~F., {Lehmann}, L.~T., {Cristofari}, P.~I., {et~al.} 2023, \mnras, 525, 2015

\bibitem[{{Donati} {et~al.}(2008){Donati}, {Morin}, {Petit}, {Delfosse}, {Forveille}, {Auri{\`e}re}, {Cabanac}, {Dintrans}, {Fares}, {Gastine}, {Jardine}, {Ligni{\`e}res}, {Paletou}, {Ramirez Velez}, \& {Th{\'e}ado}}]{Donati2008}
{Donati}, J.~F., {Morin}, J., {Petit}, P., {et~al.} 2008, \mnras, 390, 545

\bibitem[{{Espinoza} {et~al.}(2019){Espinoza}, {Kossakowski}, \& {Brahm}}]{Espinoza2019}
{Espinoza}, N., {Kossakowski}, D., \& {Brahm}, R. 2019, \mnras, 490, 2262

\bibitem[{{Foreman-Mackey} {et~al.}(2017){Foreman-Mackey}, {Agol}, {Ambikasaran}, \& {Angus}}]{Foreman-mackey2017}
{Foreman-Mackey}, D., {Agol}, E., {Ambikasaran}, S., \& {Angus}, R. 2017, \aj, 154, 220

\bibitem[{{Foreman-Mackey} {et~al.}(2013){Foreman-Mackey}, {Hogg}, {Lang}, \& {Goodman}}]{emcee}
{Foreman-Mackey}, D., {Hogg}, D.~W., {Lang}, D., \& {Goodman}, J. 2013, \pasp, 125, 306

\bibitem[{{Garcia} {et~al.}(2024){Garcia}, {Charnay}, {Rackham}, {Timmermans}, {Barkaoui}, {Bezard}, {Delrez}, {Doyon}, {Dransfield}, {Ducrot}, {Gillon}, {Lavvas}, {Lim}, {Pincon}, {Wilkinson}, \& {de Wit}}]{Garcia2024}
{Garcia}, L., {Charnay}, B., {Rackham}, B., {et~al.} 2024, {TOI-3884: A JWST Rosetta Stone for the study of M-dwarf stellar contamination}, JWST Proposal. Cycle 3, ID. \#5799

\bibitem[{{Garcia} {et~al.}(2022){Garcia}, {Timmermans}, {Pozuelos}, {Ducrot}, {Gillon}, {Delrez}, {Wells}, \& {Jehin}}]{Garcia2022}
{Garcia}, L.~J., {Timmermans}, M., {Pozuelos}, F.~J., {et~al.} 2022, \mnras, 509, 4817

\bibitem[{{Gardner} {et~al.}(2006){Gardner}, {Mather}, {Clampin}, {Doyon}, {Greenhouse}, {Hammel}, {Hutchings}, {Jakobsen}, {Lilly}, {Long}, {Lunine}, {McCaughrean}, {Mountain}, {Nella}, {Rieke}, {Rieke}, {Rix}, {Smith}, {Sonneborn}, {Stiavelli}, {Stockman}, {Windhorst}, \& {Wright}}]{Gardner2006}
{Gardner}, J.~P., {Mather}, J.~C., {Clampin}, M., {et~al.} 2006, \ssr, 123, 485

\bibitem[{{Giles} {et~al.}(2017){Giles}, {Collier Cameron}, \& {Haywood}}]{Giles2017}
{Giles}, H. A.~C., {Collier Cameron}, A., \& {Haywood}, R.~D. 2017, \mnras, 472, 1618

\bibitem[{Goodman \& Weare(2010)}]{Goodman2010}
Goodman, J. \& Weare, J. 2010, Communications in applied mathematics and computational science, 5, 65

\bibitem[{{H{\'e}brard} {et~al.}(2016){H{\'e}brard}, {Donati}, {Delfosse}, {Morin}, {Moutou}, \& {Boisse}}]{hebrard2016}
{H{\'e}brard}, {\'E}.~M., {Donati}, J.~F., {Delfosse}, X., {et~al.} 2016, \mnras, 461, 1465

\bibitem[{{Juvan} {et~al.}(2018){Juvan}, {Lendl}, {Cubillos}, {Fossati}, {Tregloan-Reed}, {Lammer}, {Guenther}, \& {Hanslmeier}}]{Juvan2018}
{Juvan}, I.~G., {Lendl}, M., {Cubillos}, P.~E., {et~al.} 2018, \aap, 610, A15

\bibitem[{{Kislyakova} {et~al.}(2018){Kislyakova}, {Fossati}, {Johnstone}, {Noack}, {L{\"u}ftinger}, {Zaitsev}, \& {Lammer}}]{Kislyakova2018}
{Kislyakova}, K.~G., {Fossati}, L., {Johnstone}, C.~P., {et~al.} 2018, \apj, 858, 105

\bibitem[{{Kislyakova} {et~al.}(2017){Kislyakova}, {Noack}, {Johnstone}, {Zaitsev}, {Fossati}, {Lammer}, {Khodachenko}, {Odert}, \& {G{\"u}del}}]{Kislyakova2017}
{Kislyakova}, K.~G., {Noack}, L., {Johnstone}, C.~P., {et~al.} 2017, Nature Astronomy, 1, 878

\bibitem[{{Kitchatinov} \& {Olemskoy}(2011)}]{Kitchatinov2011}
{Kitchatinov}, L.~L. \& {Olemskoy}, S.~V. 2011, \mnras, 411, 1059

\bibitem[{{Klein} {et~al.}(2021){Klein}, {Donati}, {Moutou}, {Delfosse}, {Bonfils}, {Martioli}, {Fouqu{\'e}}, {Cloutier}, {Artigau}, {Doyon}, {H{\'e}brard}, {Morin}, {Rameau}, {Plavchan}, \& {Gaidos}}]{Klein2021}
{Klein}, B., {Donati}, J.-F., {Moutou}, C., {et~al.} 2021, \mnras, 502, 188

\bibitem[{{Kochukhov} \& {Lavail}(2017)}]{Kochukhov2017}
{Kochukhov}, O. \& {Lavail}, A. 2017, \apjl, 835, L4

\bibitem[{{Kreidberg}(2015)}]{Kreidberg2015}
{Kreidberg}, L. 2015, \pasp, 127, 1161

\bibitem[{{K{\"u}ker} \& {R{\"u}diger}(2008)}]{kuker2008}
{K{\"u}ker}, M. \& {R{\"u}diger}, G. 2008, in Journal of Physics Conference Series, Vol. 118, Journal of Physics Conference Series (IOP), 012029

\bibitem[{{Lendl} {et~al.}(2012){Lendl}, {Anderson}, {Collier-Cameron}, {Doyle}, {Gillon}, {Hellier}, {Jehin}, {Lister}, {Maxted}, {Pepe}, {Pollacco}, {Queloz}, {Smalley}, {S{\'e}gransan}, {Smith}, {Triaud}, {Udry}, {West}, \& {Wheatley}}]{Lendl2012}
{Lendl}, M., {Anderson}, D.~R., {Collier-Cameron}, A., {et~al.} 2012, \aap, 544, A72

\bibitem[{{Libby-Roberts} {et~al.}(2023){Libby-Roberts}, {Schutte}, {Hebb}, {Kanodia}, {Ca{\~n}as}, {Stef{\'a}nsson}, {Lin}, {Mahadevan}, {Parts}, {Powers}, {Wisniewski}, {Bender}, {Cochran}, {Diddams}, {Everett}, {Gupta}, {Halverson}, {Kobulnicky}, {Kowalski}, {Larsen}, {Monson}, {Ninan}, {Parker}, {Ramsey}, {Robertson}, {Schwab}, {Swaby}, \& {Terrien}}]{Libby-Roberts2023}
{Libby-Roberts}, J.~E., {Schutte}, M., {Hebb}, L., {et~al.} 2023, \aj, 165, 249

\bibitem[{{Lightkurve Collaboration} {et~al.}(2018){Lightkurve Collaboration}, {Cardoso}, {Hedges}, {Gully-Santiago}, {Saunders}, {Cody}, {Barclay}, {Hall}, {Sagear}, {Turtelboom}, {Zhang}, {Tzanidakis}, {Mighell}, {Coughlin}, {Bell}, {Berta-Thompson}, {Williams}, {Dotson}, \& {Barentsen}}]{lightcurve2018}
{Lightkurve Collaboration}, {Cardoso}, J. V. d.~M., {Hedges}, C., {et~al.} 2018, {Lightkurve: Kepler and TESS time series analysis in Python}, Astrophysics Source Code Library, record ascl:1812.013

\bibitem[{{Lim} {et~al.}(2023){Lim}, {Benneke}, {Doyon}, {MacDonald}, {Piaulet}, {Artigau}, {Coulombe}, {Radica}, {L'Heureux}, {Albert}, {Rackham}, {de Wit}, {Salhi}, {Roy}, {Flagg}, {Fournier-Tondreau}, {Taylor}, {Cook}, {Lafreni{\`e}re}, {Cowan}, {Kaltenegger}, {Rowe}, {Espinoza}, {Dang}, \& {Darveau-Bernier}}]{Lim2023}
{Lim}, O., {Benneke}, B., {Doyon}, R., {et~al.} 2023, \apjl, 955, L22

\bibitem[{{Lomb}(1976)}]{lomb}
{Lomb}, N.~R. 1976, \apss, 39, 447

\bibitem[{{Moran} {et~al.}(2023){Moran}, {Stevenson}, {Sing}, {MacDonald}, {Kirk}, {Lustig-Yaeger}, {Peacock}, {Mayorga}, {Bennett}, {L{\'o}pez-Morales}, {May}, {Rustamkulov}, {Valenti}, {Adams Redai}, {Alam}, {Batalha}, {Fu}, {Gonzalez-Quiles}, {Highland}, {Kruse}, {Lothringer}, {Ortiz Ceballos}, {Sotzen}, \& {Wakeford}}]{Moran2023}
{Moran}, S.~E., {Stevenson}, K.~B., {Sing}, D.~K., {et~al.} 2023, \apjl, 948, L11

\bibitem[{{Mori} {et~al.}(2025){Mori}, {Fukui}, {Hirano}, {Narita}, {Livingston}, {Barkaoui}, {Collins}, {de Leon}, {Ikuta}, {Kawai}, {Schwarz}, {Shporer}, \& {Srdoc}}]{Mori2025}
{Mori}, M., {Fukui}, A., {Hirano}, T., {et~al.} 2025, arXiv e-prints, arXiv:2506.06445

\bibitem[{{Morin} {et~al.}(2008){Morin}, {Donati}, {Petit}, {Delfosse}, {Forveille}, {Albert}, {Auri{\`e}re}, {Cabanac}, {Dintrans}, {Fares}, {Gastine}, {Jardine}, {Ligni{\`e}res}, {Paletou}, {Ramirez Velez}, \& {Th{\'e}ado}}]{Morin2008}
{Morin}, J., {Donati}, J.~F., {Petit}, P., {et~al.} 2008, \mnras, 390, 567

\bibitem[{{Morin} {et~al.}(2010){Morin}, {Donati}, {Petit}, {Delfosse}, {Forveille}, \& {Jardine}}]{Morin2010}
{Morin}, J., {Donati}, J.~F., {Petit}, P., {et~al.} 2010, \mnras, 407, 2269

\bibitem[{{Mo{\v{c}}nik} {et~al.}(2016){Mo{\v{c}}nik}, {Clark}, {Anderson}, {Hellier}, \& {Brown}}]{Mocnik2016}
{Mo{\v{c}}nik}, T., {Clark}, B.~J.~M., {Anderson}, D.~R., {Hellier}, C., \& {Brown}, D.~J.~A. 2016, \aj, 151, 150

\bibitem[{{Murray} {et~al.}(2024){Murray}, {Berta-Thompson}, {Canas}, {Feinstein}, {Hebb}, {Kanodia}, {Libby-Roberts}, {Mahadevan}, {Morley}, {Ninan}, {Stefansson}, \& {Welbanks}}]{Murray2024}
{Murray}, C.~A., {Berta-Thompson}, Z.~K., {Canas}, C., {et~al.} 2024, {Shining a Spot-light on the Atmosphere of a Giant Planet around a Cool Star}, JWST Proposal. Cycle 3, ID. \#5863

\bibitem[{{Namekata} {et~al.}(2020){Namekata}, {Davenport}, {Morris}, {Hawley}, {Maehara}, {Notsu}, {Toriumi}, {Ikuta}, {Notsu}, {Honda}, {Nogami}, \& {Shibata}}]{Namekata2020}
{Namekata}, K., {Davenport}, J. R.~A., {Morris}, B.~M., {et~al.} 2020, \apj, 891, 103

\bibitem[{{Namekata} {et~al.}(2019){Namekata}, {Maehara}, {Notsu}, {Toriumi}, {Hayakawa}, {Ikuta}, {Notsu}, {Honda}, {Nogami}, \& {Shibata}}]{Namekata2019}
{Namekata}, K., {Maehara}, H., {Notsu}, Y., {et~al.} 2019, \apj, 871, 187

\bibitem[{{Pont} {et~al.}(2008){Pont}, {Knutson}, {Gilliland}, {Moutou}, \& {Charbonneau}}]{Pont2008}
{Pont}, F., {Knutson}, H., {Gilliland}, R.~L., {Moutou}, C., \& {Charbonneau}, D. 2008, \mnras, 385, 109

\bibitem[{{Rackham} {et~al.}(2018){Rackham}, {Apai}, \& {Giampapa}}]{Rackham2018}
{Rackham}, B.~V., {Apai}, D., \& {Giampapa}, M.~S. 2018, \apj, 853, 122

\bibitem[{{Ricker} {et~al.}(2015){Ricker}, {Winn}, {Vanderspek}, {Latham}, {Bakos}, {Bean}, {Berta-Thompson}, {Brown}, {Buchhave}, {Butler}, {Butler}, {Chaplin}, {Charbonneau}, {Christensen-Dalsgaard}, {Clampin}, {Deming}, {Doty}, {De Lee}, {Dressing}, {Dunham}, {Endl}, {Fressin}, {Ge}, {Henning}, {Holman}, {Howard}, {Ida}, {Jenkins}, {Jernigan}, {Johnson}, {Kaltenegger}, {Kawai}, {Kjeldsen}, {Laughlin}, {Levine}, {Lin}, {Lissauer}, {MacQueen}, {Marcy}, {McCullough}, {Morton}, {Narita}, {Paegert}, {Palle}, {Pepe}, {Pepper}, {Quirrenbach}, {Rinehart}, {Sasselov}, {Sato}, {Seager}, {Sozzetti}, {Stassun}, {Sullivan}, {Szentgyorgyi}, {Torres}, {Udry}, \& {Villasenor}}]{Ricker2015}
{Ricker}, G.~R., {Winn}, J.~N., {Vanderspek}, R., {et~al.} 2015, Journal of Astronomical Telescopes, Instruments, and Systems, 1, 014003

\bibitem[{{Roettenbacher} {et~al.}(2016){Roettenbacher}, {Monnier}, {Korhonen}, {Aarnio}, {Baron}, {Che}, {Harmon}, {K{\H{o}}v{\'a}ri}, {Kraus}, {Schaefer}, {Torres}, {Zhao}, {Ten Brummelaar}, {Sturmann}, \& {Sturmann}}]{Rottenbacher2016}
{Roettenbacher}, R.~M., {Monnier}, J.~D., {Korhonen}, H., {et~al.} 2016, \nat, 533, 217

\bibitem[{{Sanchis-Ojeda} \& {Winn}(2011)}]{Sanchis-Ojeda2011}
{Sanchis-Ojeda}, R. \& {Winn}, J.~N. 2011, \apj, 743, 61

\bibitem[{{Sanchis-Ojeda} {et~al.}(2013){Sanchis-Ojeda}, {Winn}, {Marcy}, {Howard}, {Isaacson}, {Johnson}, {Torres}, {Albrecht}, {Campante}, {Chaplin}, {Davies}, {Lund}, {Carter}, {Dawson}, {Buchhave}, {Everett}, {Fischer}, {Geary}, {Gilliland}, {Horch}, {Howell}, \& {Latham}}]{Sanchis-Ojeda2013}
{Sanchis-Ojeda}, R., {Winn}, J.~N., {Marcy}, G.~W., {et~al.} 2013, \apj, 775, 54

\bibitem[{{Scargle}(1982)}]{scargle}
{Scargle}, J.~D. 1982, \apj, 263, 835

\bibitem[{{See} {et~al.}(2025){See}, {Amard}, {Bellotti}, {Saikia}, {Brown}, {Donati}, {Fares}, {Finley}, {Folsom}, {H{\'e}brard}, {Jardine}, {Jeffers}, {Klein}, {Lehmann}, {Marsden}, {Matt}, {Mengel}, {Morin}, {Petit}, {Smith}, {Vidotto}, \& {Waite}}]{See2025}
{See}, V., {Amard}, L., {Bellotti}, S., {et~al.} 2025, \mnras [\eprint[arXiv]{2507.16986}]

\bibitem[{{Silva}(2003)}]{Silva2003}
{Silva}, A. V.~R. 2003, \apjl, 585, L147

\bibitem[{{Speagle}(2020)}]{Speagle2020}
{Speagle}, J.~S. 2020, \mnras, 493, 3132

\bibitem[{{Strassmeier}(2009)}]{Strassmeier2009}
{Strassmeier}, K.~G. 2009, \aapr, 17, 251

\bibitem[{{Szab{\'o}} {et~al.}(2021){Szab{\'o}}, {Gandolfi}, {Brandeker}, {Csizmadia}, {Garai}, {Billot}, {Broeg}, {Ehrenreich}, {Fortier}, {Fossati}, {Hoyer}, {Kiss}, {Lecavelier des Etangs}, {Maxted}, {Ribas}, {Alibert}, {Alonso}, {Anglada Escud{\'e}}, {B{\'a}rczy}, {Barros}, {Barrado}, {Baumjohann}, {Beck}, {Beck}, {Bekkelien}, {Bonfils}, {Benz}, {Borsato}, {Busch}, {Cabrera}, {Charnoz}, {Collier Cameron}, {Van Damme}, {Davies}, {Delrez}, {Deleuil}, {Demangeon}, {Demory}, {Erikson}, {Fridlund}, {Futyan}, {Garc{\'\i}a Mu{\~n}oz}, {Gillon}, {Guedel}, {Guterman}, {Heng}, {Isaak}, {Lacedelli}, {Laskar}, {Lendl}, {Lovis}, {Luntzer}, {Magrin}, {Nascimbeni}, {Olofsson}, {Osborn}, {Ottensamer}, {Pagano}, {Pall{\'e}}, {Peter}, {Piazza}, {Piotto}, {Pollacco}, {Queloz}, {Ragazzoni}, {Rando}, {Rauer}, {Santos}, {Scandariato}, {S{\'e}gransan}, {Serrano}, {Sicilia}, {Simon}, {Smith}, {Sousa}, {Steller}, {Thomas}, {Udry}, {Van Grootel}, {Walton}, \& {Wilson}}]{Szabo2021}
{Szab{\'o}}, G.~M., {Gandolfi}, D., {Brandeker}, A., {et~al.} 2021, \aap, 654, A159

\bibitem[{{Tamburo} {et~al.}(2025){Tamburo}, {Yee}, {Garc{\'\i}a-Mej{\'\i}a}, {Charbonneau}, {Bieryla}, {Collins}, \& {Shporer}}]{Tamburo2025}
{Tamburo}, P., {Yee}, S.~W., {Garc{\'\i}a-Mej{\'\i}a}, J., {et~al.} 2025, arXiv e-prints, arXiv:2506.11998

\bibitem[{{Vidotto} {et~al.}(2014){Vidotto}, {Gregory}, {Jardine}, {Donati}, {Petit}, {Morin}, {Folsom}, {Bouvier}, {Cameron}, {Hussain}, {Marsden}, {Waite}, {Fares}, {Jeffers}, \& {do Nascimento}}]{Vidotto2014}
{Vidotto}, A.~A., {Gregory}, S.~G., {Jardine}, M., {et~al.} 2014, \mnras, 441, 2361

\bibitem[{{Wang} \& {Espinoza}(2024)}]{Wang2024}
{Wang}, G. \& {Espinoza}, N. 2024, \aj, 167, 1

\bibitem[{{Wheatley} {et~al.}(2018){Wheatley}, {West}, {Goad}, {Jenkins}, {Pollacco}, {Queloz}, {Rauer}, {Udry}, {Watson}, {Chazelas}, {Eigm{\"u}ller}, {Lambert}, {Genolet}, {McCormac}, {Walker}, {Armstrong}, {Bayliss}, {Bento}, {Bouchy}, {Burleigh}, {Cabrera}, {Casewell}, {Chaushev}, {Chote}, {Csizmadia}, {Erikson}, {Faedi}, {Foxell}, {G{\"a}nsicke}, {Gillen}, {Grange}, {G{\"u}nther}, {Hodgkin}, {Jackman}, {Jord{\'a}n}, {Louden}, {Metrailler}, {Moyano}, {Nielsen}, {Osborn}, {Poppenhaeger}, {Raddi}, {Raynard}, {Smith}, {Soto}, \& {Titz-Weider}}]{Wheatley2018}
{Wheatley}, P.~J., {West}, R.~G., {Goad}, M.~R., {et~al.} 2018, \mnras, 475, 4476

\bibitem[{{Willamo} {et~al.}(2022){Willamo}, {Lehtinen}, {Hackman}, {K{\"a}pyl{\"a}}, {Kochukhov}, {Jeffers}, {Korhonen}, \& {Marsden}}]{Willamo2022}
{Willamo}, T., {Lehtinen}, J.~J., {Hackman}, T., {et~al.} 2022, \aap, 659, A71

\end{thebibliography}

\begin{appendix}
\onecolumn
\section{Additional figures}

\begin{figure}[htp]
   \centering
    \includegraphics[width=\textwidth]{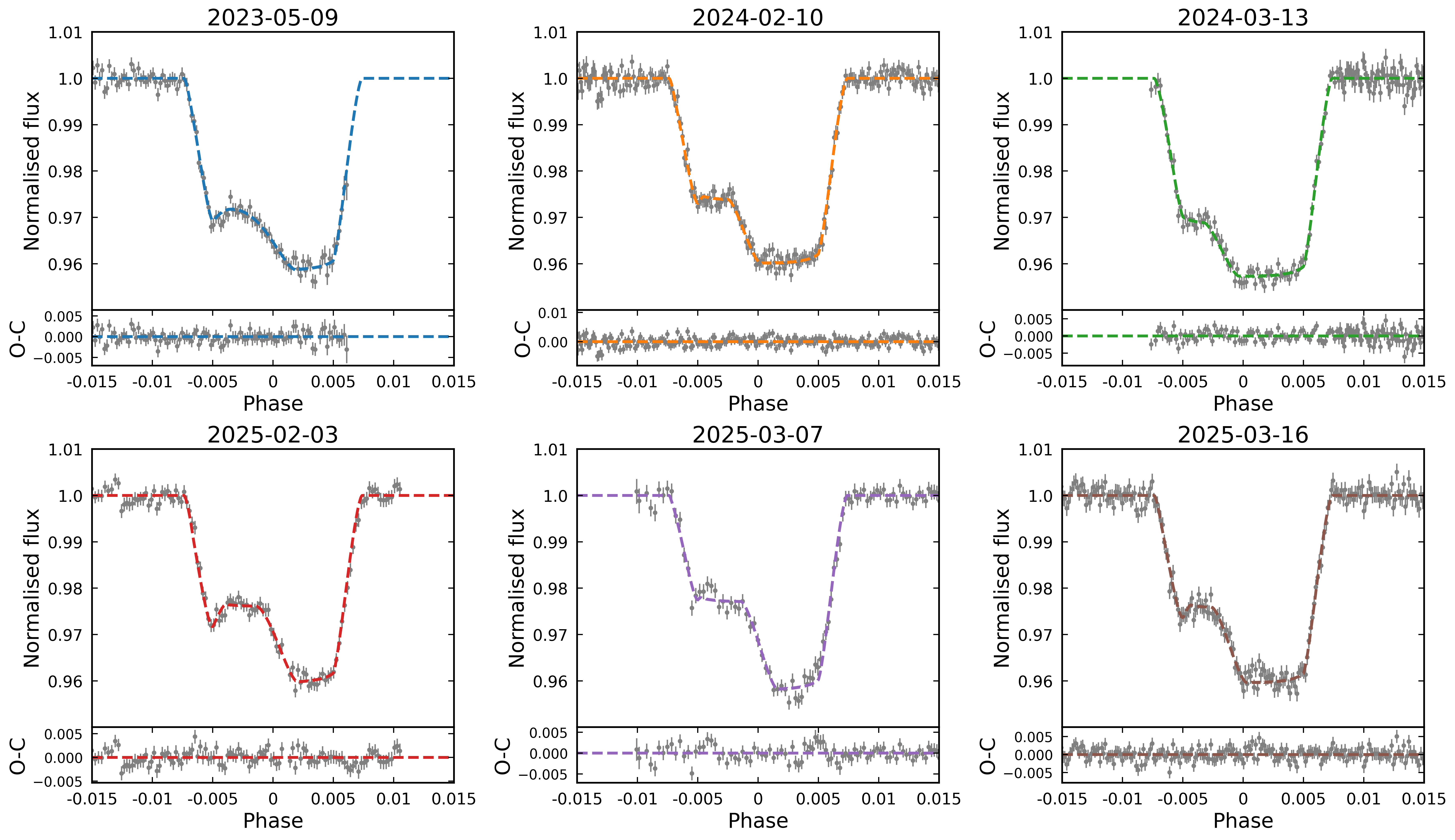}
      \caption{Same as Fig. \ref{fig:lightcurves_approach_2} but with best-fit \pytranspot models obtained using the first (or spot evolution) approach. }
         \label{fig:lightcurves_approach_1}
\end{figure}


\begin{figure}[b]
  \centering
  \includegraphics[width=1\textwidth]{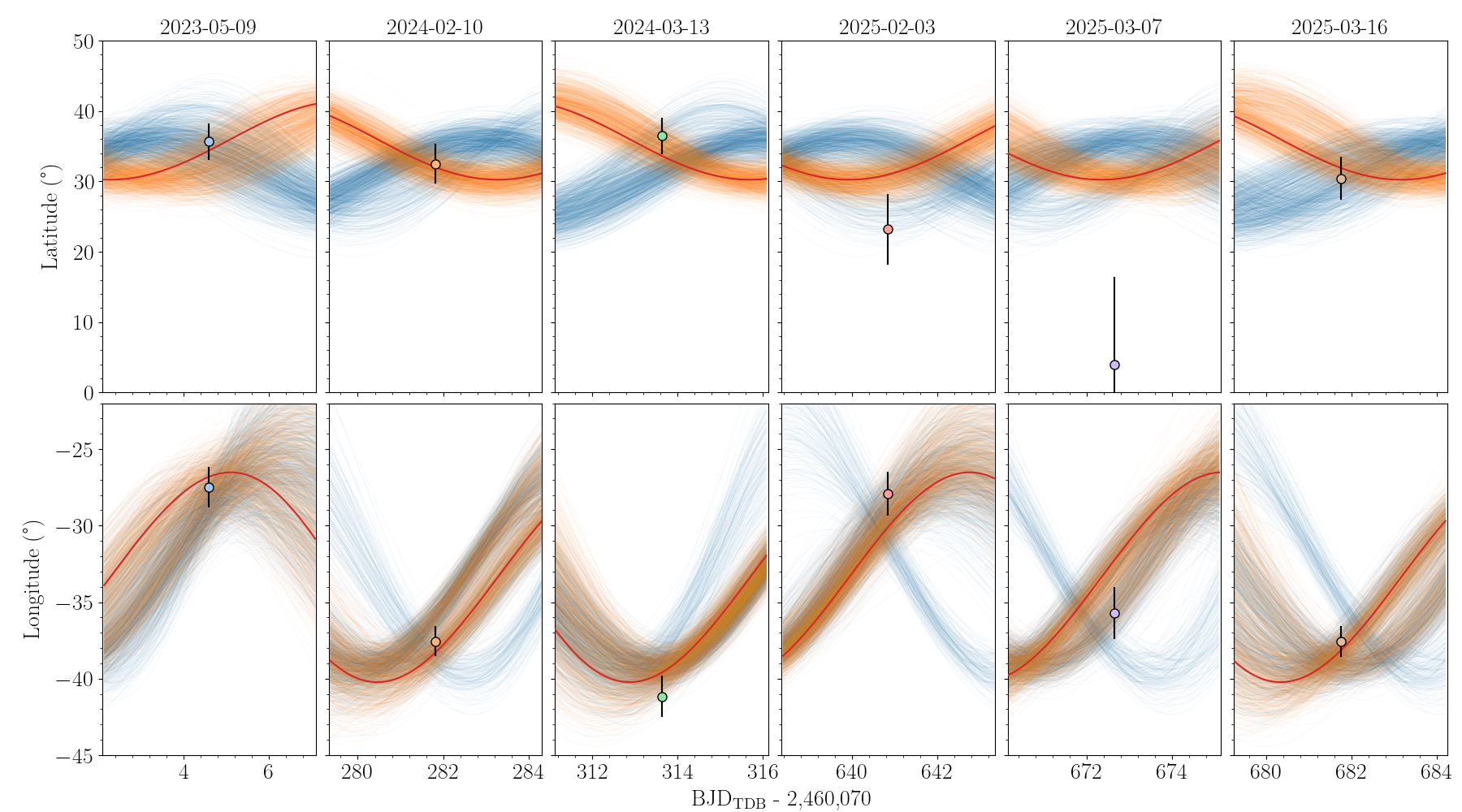}
  \caption{Modelling of the spot center’s position in latitude and longitude from the second approach analysis (i.e. the spin-spot misalignment approach) for each transit observation using the spin-spot misalignment model (Sect.~\ref{subsection:obliquity}). A thousand random samples from the posterior distribution for each direction, CW (light blue) and CCW (light orange, are shown. The best-fit model for a rotational period of $\sim10.8$~days in the CCW direction is represented by an orange line. } \label{figure:model}
\end{figure}

\begin{figure*}
  \centering
  \includegraphics[width=0.49\textwidth]{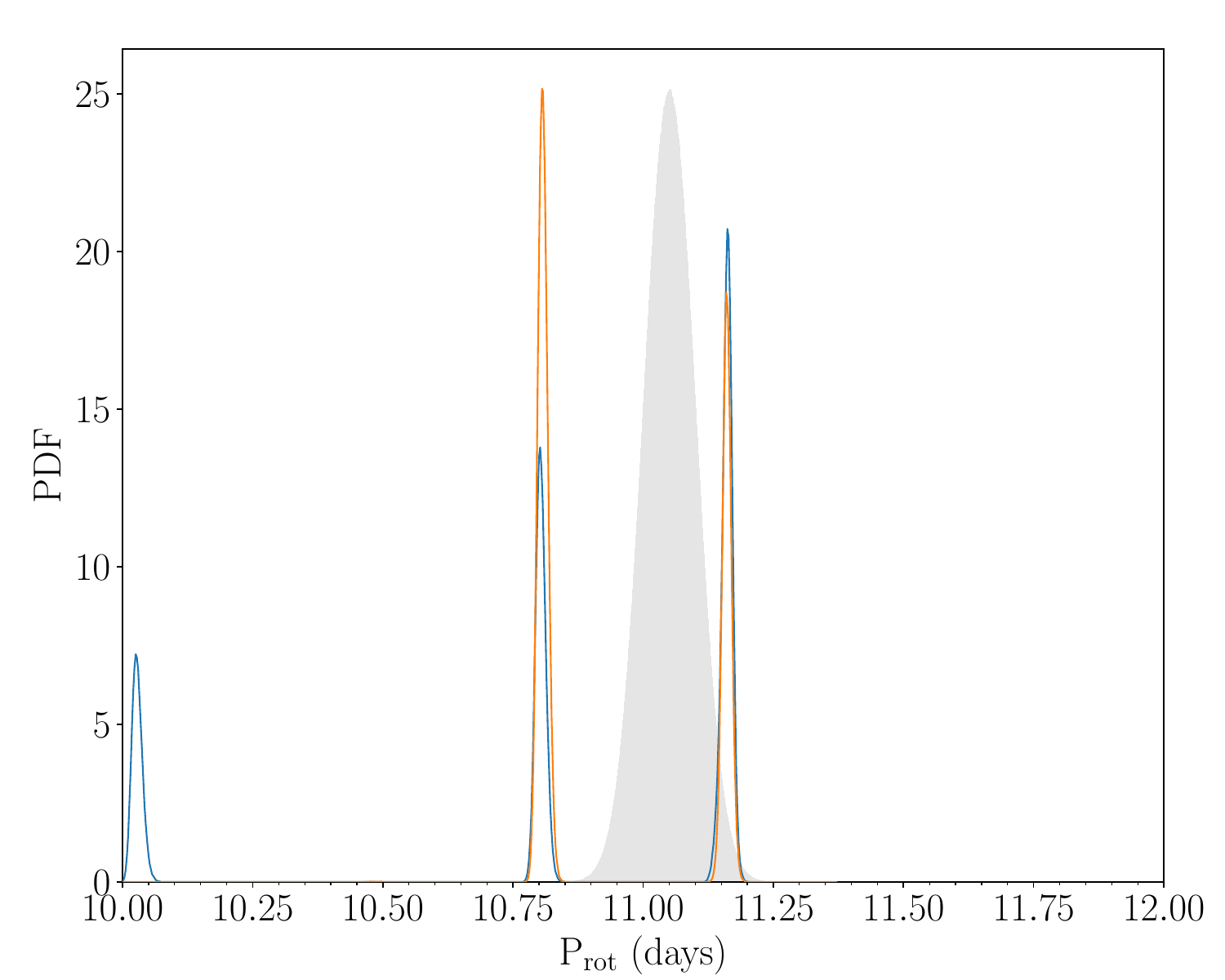}
  \caption{Posterior distribution of the rotational period from the spin-spot misalignment model (Sect.~\ref{subsection:obliquity}), shown in blue for the CW direction and in orange for the CCW direction. The gray distribution in the background represents the rotational period determined by \citet{Mori2025}.} \label{figure:Prot}
\end{figure*}

\begin{figure}[htp]
   \centering
    \includegraphics[width=\textwidth]{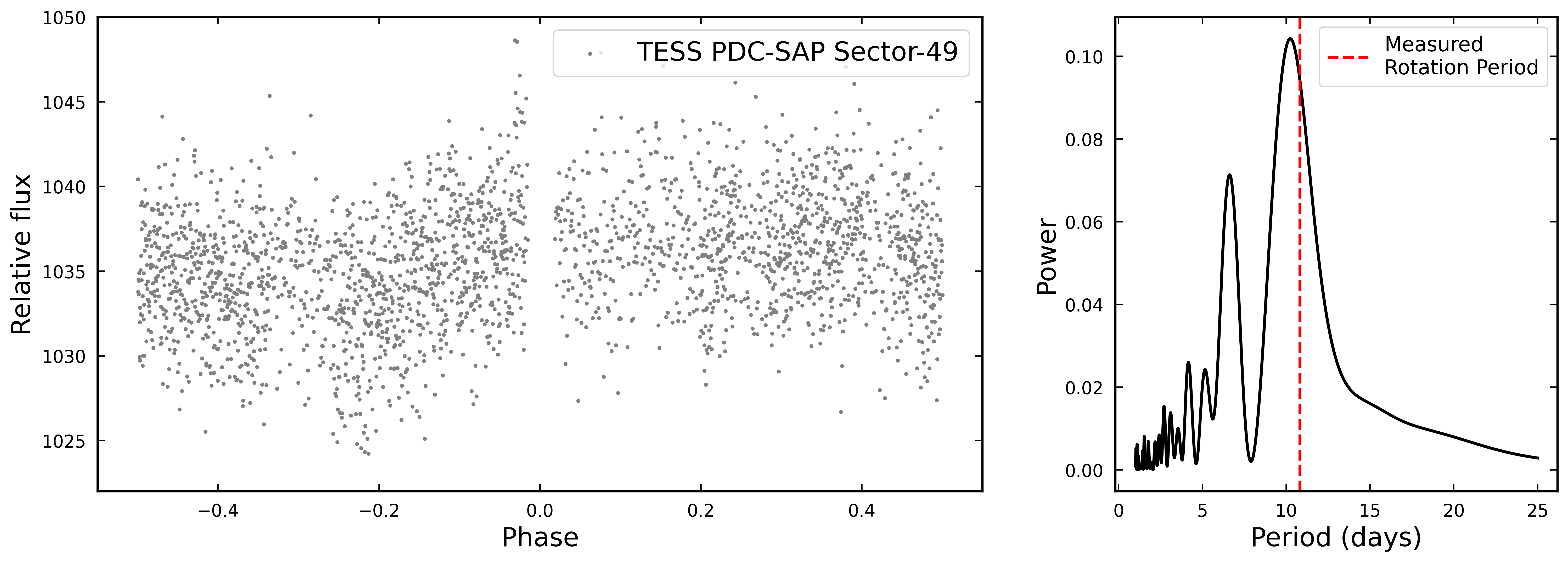}
      \caption{Phase folded light curves of TOI-3884 from TESS sector-49. The right panel displays the Lomb-Scargle periodograms with the measured rotation period from the spin-spot misalignment analysis.}
         \label{fig:tess_sector_49}
\end{figure}

\section{Posteriors of spot parameters}
\begin{table*}[h!]
    \tiny
    \renewcommand{\arraystretch}{1.25}
    \setlength{\tabcolsep}{10pt}
\centering
\caption{Posteriors of spot parameters using the first (or spot evolution) approach}\label{table:spot_parameters}
\begin{tabular}{lccclllll}
\hline
Parameter & Units & Prior & \multicolumn{6}{c}{Posterior median and $1\sigma$}\\
 & & & 2023-05-09& 2024-02-10&2024-03-13& 2025-02-03& 2025-03-07&2025-03-16\\
\hline
Latitude& [\degree] & $U(-90, 90)$ & $41 \pm 11$   & 87$^{+10}_{-8}$ & 89$^{+21}_{-19}$& 67$^{+21}_{-12}$& 61$^{+13}_{-9}$&83$^{+13}_{-11}$\\
Longitude & [\degree] & $U(-180, 180)$ & $-32 \pm 3$   & -39 $\pm$ 1 & -72$^{+16}_{-12}$ & -28.5 $\pm$ 1.5& -41$^{+5}_{-23}$& -37 $\pm$ 1\\
Size& [\degree] & $U(0, 90)$ & $50 \pm 7$   & 30$^{+2}_{-1}$&  59$^{+11}_{-15}$& 38$^{+7}_{-4}$& 50$^{+18}_{-8}$& 28$^{+2}_{-1}$\\
Contrast&  & $U(0, 1)$ & $0.60^{+0.08}_{-0.18}$   & 0.66 $\pm$ 0.01& 0.7 $\pm$ 0.01& 0.59 $\pm$ 0.02& 0.55 $\pm$ 0.02& 0.62 $\pm$ 0.01\\
\hline
\end{tabular}
\end{table*}

\begin{table*}[h!]
    \tiny
    \renewcommand{\arraystretch}{1.25}
    \setlength{\tabcolsep}{10pt}
\centering
\caption{Posteriors of spot parameters using the second (or spin-spot misalignment) approach}\label{table:spot_parameters_approach2}
\begin{tabular}{lccclllll}
\hline
Parameter & Units & Prior & \multicolumn{6}{c}{Posterior median and $1\sigma$}\\
 & & & 2023-05-09& 2024-02-10&2024-03-13& 2025-02-03& 2025-03-07&2025-03-16\\
\hline
Latitude& [\degree] & $U(-90, 90)$ & $54.1 \pm 2.5$   & $57 \pm 3$ & 53.5$^{+2.4}_{-2.8}$& 66$^{+4.7}_{-4.4}$& 87$^{+13}_{-14}$&$59 \pm 3$\\
Longitude & [\degree] & $U(-180, 180)$ & $-27 \pm 1$   & -37 $\pm$ 1 & -41.2$^{+1.4}_{-1.5}$ & -28 $\pm$ 1.4& -35.7$^{+1.6}_{-1.8}$& -38 $\pm$ 1\\
\hline
\end{tabular}
\tablefoot{The combined spot size and contrast are $38 \pm 2\degree$ and $0.58 \pm 0.01$}
\end{table*}

\begin{figure}
  \centering
  \includegraphics[width=0.49\textwidth]{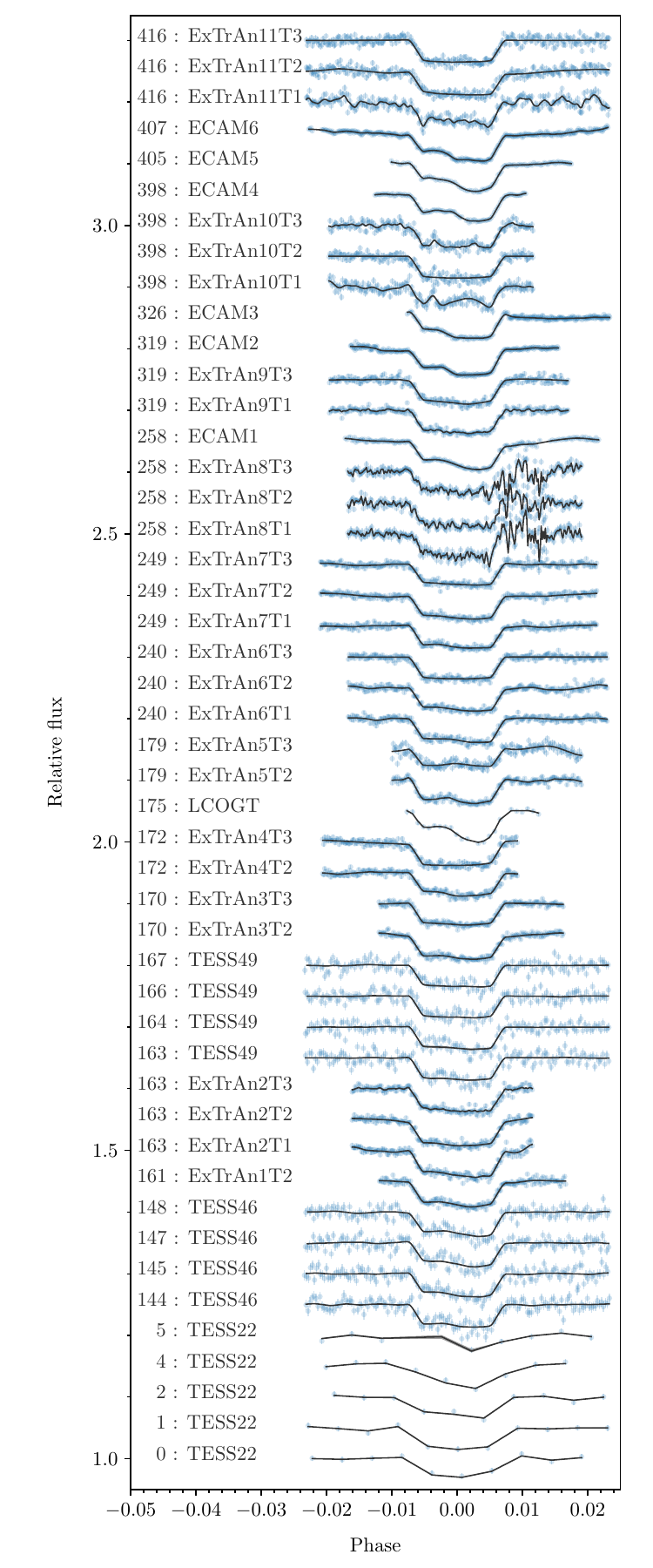}
  \caption{TESS, ExTrA, LCOGT, and ECAM transits (offset for clarity) modelled with \juliet. The blue symbols with error bars are the data, and the black line is the \juliet posterior median model. Each transit is labelled with the epoch relative to the first transit observed by TESS, the sector for transits observed with TESS, and the night and the telescope for transits observed with ExTrA.} \label{fig:juliet}
\end{figure}

\end{appendix}
\end{document}